# Multi-Source Domain Transfer Learning for Accurate Property Prediction in Two-Dimensional Materials

Huiyang Zhang[1], Xinyu Chen[1], Qionghua Zhou[1,*], and Jinlan Wang[1,*]

[1]Key Laboratory of Quantum Materials and Devices of Ministry of Education, School of Physics, Southeast University, Nanjing 211189, China

[*]E-mail: qh.zhou@seu.edu.cn (Q. Z.); jlwang@seu.edu.cn (J. W.)

## Abstract

Machine learning has revolutionized materials discovery, but data scarcity remains a critical bottleneck for complex functional properties. As emerging systems, two-dimensional (2D) materials possess limited overall data volumes. Evaluating their diverse functional properties requires time-consuming simulations, hindering unified high-throughput screening. Furthermore, restrictions in known structural prototypes lead to highly fragmented data distributions. To address these challenges, we propose a multi-source domain transfer learning framework to extract generalizable and complementary knowledge from diverse crystalline systems. To mitigate data scarcity, the framework employs a shared feature extractor that integrates adversarial transfer learning with maximum mean discrepancy, mapping crystal structures into a domain-invariant latent space while preserving underlying physical correlations. To resolve distribution fragmentation, a sample-adaptive weighted ensemble strategy is subsequently utilized to dynamically aggregate predictions from multiple source domains. Relying solely on crystal structures, the framework predicts 2D carrier mobilities with an $R^2$ score exceeding 0.90. The framework successfully screened 55 novel high-mobility 2D semiconductors, which were validated via first-principles electron-phonon coupling analysis, confirming their exceptional transport properties and stability. This work can potentially accelerate machine learning-assisted materials design and discovery with less data restriction.

## Introduction

Machine learning (ML) has fundamentally reshaped the landscape of materials informatics[1,2], establishing high-fidelity surrogate models as a standard paradigm for predicting foundational properties such as formation energy[3], structural stability[4], and bandgaps[5]. While these data-driven approaches reduce reliance on costly experimental trials, their application to complex functional properties remains limited. Specifically, accurate evaluation of carrier transport properties requires computationally intensive first-principles simulations of electron–phonon interactions, severely restricting the availability of high-fidelity datasets for large-scale screening. Consequently, a fundamental bottleneck persists, as the efficacy of ML models intrinsically depends on extensive and high-quality datasets[6–8], whereas data for complex functional properties are often highly inhomogeneous and scarce. Although major DFT databases provide millions of structures and basic energetic properties[9–11], high-fidelity functional properties remain scarce because of the prohibitive cost of advanced simulations[12]. Active learning partially alleviates this issue but still depends on expensive iterative calculations and suffers from sampling bias. Consequently, approaches that transfer knowledge from related systems offer a promising alternative.

Transfer learning (TL) serves as a promising paradigm that effectively circumvents data scarcity issues by transferring physics knowledge from data-rich source domains to data-limited target domains[13,14]. For instance, Grunert et al. demonstrated that a TL-based approach could achieve predictive accuracy for optoelectronic properties using only 300 samples, comparable to direct training with 3,000 samples[15]. Similarly, Chen et al. developed a hybrid framework integrating adversarial training and expert knowledge to extract shared features from bulk effective mass data, enabling accurate predictions of 2D material carrier mobility through cross-material and cross-property knowledge transfer[16]. Despite these successes, conventional transfer learning applications in materials science predominantly rely on a single source domain[17–21]. This single-source paradigm usually depends on expert intuition and empirical trial-and-error for source domain selection, which may not only introduce subjective bias but also risk suboptimal predictive performance if an inappropriate source domain is selected.

While researchers typically seek a single source domain with high physical similarity, such ideal systems are rare and often provide only partial information regarding the underlying physics. For example, a source might have a similar crystal structure but have very different electronic behavior[22,23], capturing only a fragmented subset of the required physical rules. This incomplete knowledge representation renders the transfer process fragile and unreliable[24,25]. Moreover, TL from a source task dissimilar to the downstream task can even lead to worse performance than training a model on the downstream task from scratch, a phenomenon known as negative transfer[26]. Consequently, relying on a single source domain remains an unsustainable bottleneck for robust cross-material knowledge transfer.

To address the partial information bottleneck of single-source transfer, leveraging multiple source domains across distinct material properties and dimensions offers a compelling pathway, as fundamental energetic and transport characteristics inherently provide complementary physical insights into identical atomic structures. The prediction of carrier mobility in 2D semiconductors exemplifies both the necessity and the challenge of this multi-source strategy. While evaluating 2D mobility suffers from high computational demands[27], abundant data already exists for 2D formation energies and 3D bulk counterparts. However, effective utilization of these heterogeneous data poses a great challenge in jointly aligning source domains from diverse material systems with the target domain[12]. Without such simultaneous alignment, a naive aggregation of these diverse sources inevitably induces negative transfer, potentially neutralizing the benefits of multi-source data abundance.

In this work, we develop a multi-source transfer learning framework to solve these problems. Our approach goes beyond simple one-to-one mapping by combining knowledge from many different source domains. These sources include related material properties and crystal structures. Instead of using a fixed rule, our framework uses an adaptive weighting system to automatically find and use the most helpful physical features for each material. To ensure the model is accurate, we use the latest electron-phonon coupling (EPC) methods to create a high-fidelity dataset for 2D carrier mobility. This high-quality data allows the model to learn the real physics of transport even when data is scarce. Our strategy effectively avoids the risk of negative transfer and works well with very little data. The model can predict 2D carrier mobility by using only crystal structure files as the input. Using this framework, we successfully identified 55 new 2D materials with high mobility above 1,000 $cm^2V^{-1}s^{-1}$. This work provides a fast and reliable way to discover next-generation 2D semiconductors for electronic devices.

## 2. Results and Discussion

### 2.1. Multi-source domain transfer learning framework

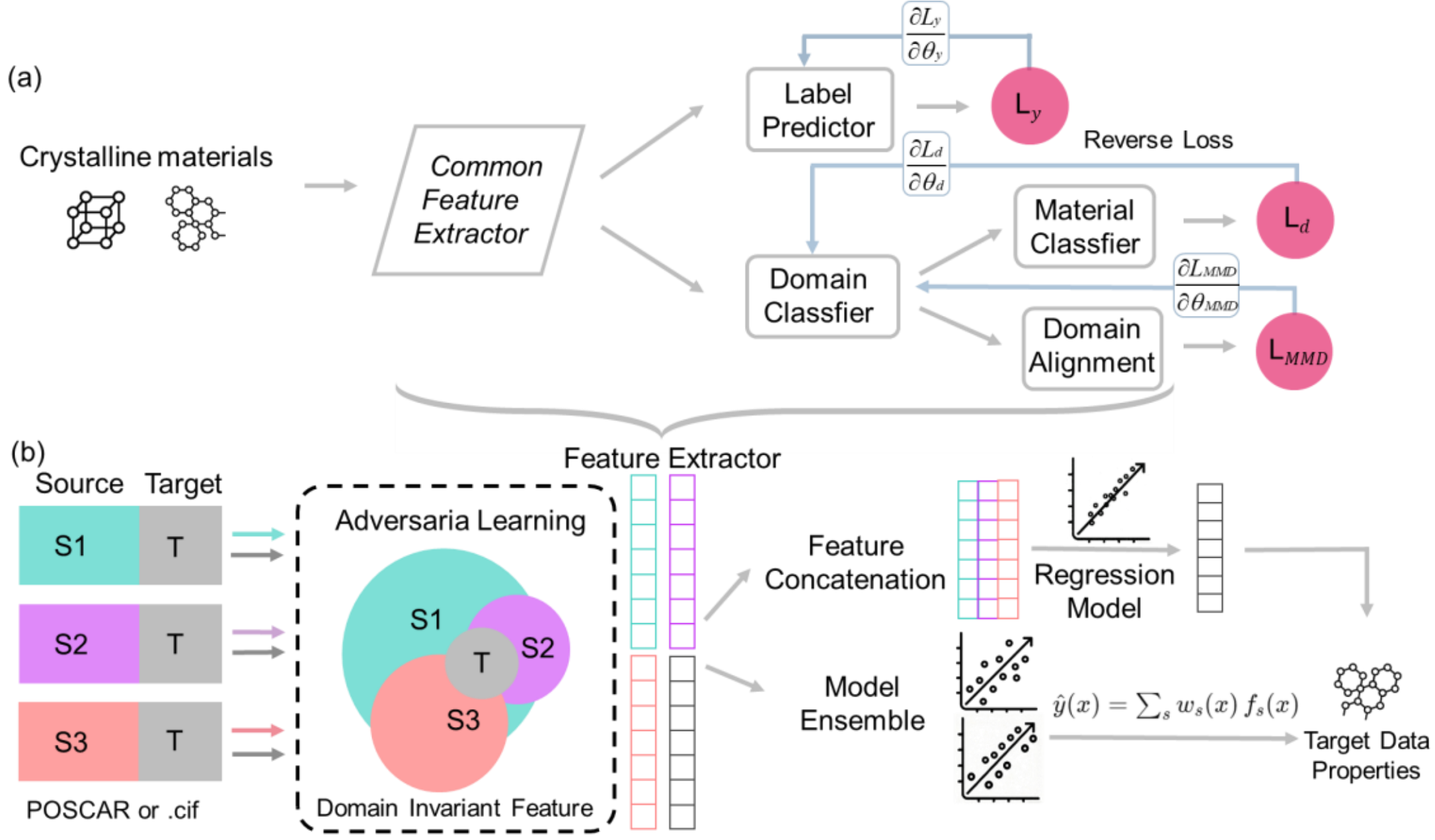

**Figure 1. Schematic of the Multi-source Domain Transfer Learning Framework**. (**a**) Model Architecture. Crystalline structures and compositions are encoded into high-dimensional vectors via MAGPIE and processed by a Common Feature Extractor, which simultaneously feeds into three functional branches: a Label Predictor, a Material Classifier, and a Domain Alignment module to extract cross-domain invariant representations. **(b)** Multi-source integration strategies. Comparison of two knowledge fusion paradigms: (i) Feature Concatenation, which merges latent vectors for joint regression, and (ii) Model Ensemble, which combines source-specific regressors via dynamic, sample-adaptive weighting to forecast target attributes.

The proposed multi-source transfer learning framework integrates physical knowledge from different crystal systems to improve property prediction for 2D materials. As shown in Figure 1, the architecture has two main parts that work together. These parts include a domain alignment module and a multi-source feature integration pipeline. This structure allows the model to learn from both property-related and structure-related data while managing the large differences between bulk and 2D materials.

The first part of our framework uses transfer learning to extract shared features from both bulk materials and 2D materials as shown in Figure 1a. We start with multiple source domains $\{S_i\}$ where each domain represents a different material property. Together with a target domain T, all samples are processed by a shared feature extractor to learn a unified latent representation. This extractor turns the initial crystal data into a low-dimensional vector by using the materials-agnostic platform for informatics and exploration (MAGPIE) platform[28]. The model then follows three cooperative learning goals to refine these features with Adversarial Transfer Learning(ATL). First, a label predictor is trained on the source domains to perform supervised regression. This step ensures that the learned features contain the right information about material properties. Second, we introduce a domain classifier to distinguish between source and target domains. By training the model with a reverse loss, the feature extractor learns to ignore details that are only found in one specific domain. Third, an maximum mean discrepancy (MMD) loss is added to measure the difference between the source and target feature distributions. By minimizing this loss, we directly reduce the statistical distance between the two domains in the hidden space. During the training process, the feature extractor is optimized by all three losses at the same time. This results in features that are both informative for property prediction and consistent across different material systems. By removing domain-specific noise, the framework keeps only the universal physical descriptors that are needed for successful cross-material transfer.

The second part of our framework consolidates multiple source domains (S1, S2, …, $S_M$) and the target domain (T), as schematically illustrated with three representative sources ($S_1$, $S_2$, $S_3$) in Figure 1b. This setup allows the model to transfer knowledge within a single unified space. Before the domains are aligned, samples from different sources usually stay in separate groups. The target data might only overlap with some sources or be completely different from them. This happens because different physical properties follow different rules. By using domain alignment and MMD, the feature extractor learns to find features that are the same across domains but still keep the important physical information. This reduces the difference between the target and the most relevant sources while ignoring the sources that do not match well. We tested two ways to combine this knowledge to find the best strategy. The first way is the Feature Concatenation Pathway. In this method, we join the features from all sources into one long vector and feed it into a single model. However, this often gives poor results. When the target data is very small, a very long feature vector makes the model too complex and leads to overfitting. It also introduces noise from irrelevant sources that makes the predictions less stable. To solve these problems, we created the Gated Model Ensemble Pathway. Instead of mixing the raw features, we build a separate expert model $f_s(x)$ for each source. We then use an adaptive strategy to calculate a set of weights $\omega_s(x)$ based on the physical features of the input sample $x$. The final prediction is a weighted sum $\hat{y}(x) = \sum_s \omega_s(x) f_s(x)$. By combining the models instead of the features, we avoid the problem of having too many dimensions. More importantly, this method lets the model choose the most useful knowledge and stop negative transfer from sources that do not relate to the target. This ensures the model works well for discovering 2D materials even when we have very little data.

### 2.2. Performance of the model

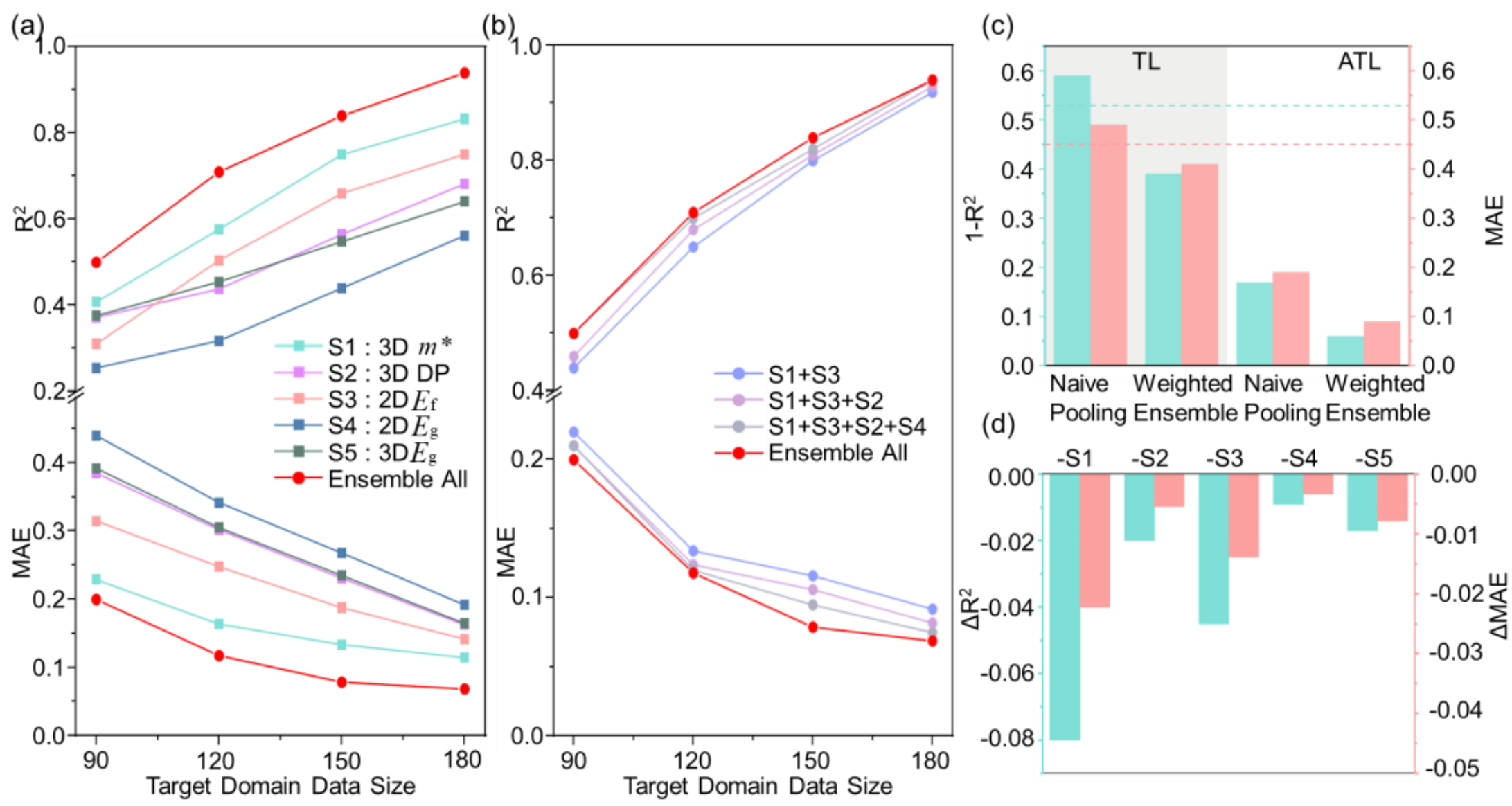

**Figure 2. Performance Comparison of Multi-Source Domain Transfer Learning in Target Material Property Prediction.** (**a**) This figure demonstrates the prediction performance of five single-source domains (S1–S5) and their integrated model across varying sample sizes in target domains. (**b**) The performance of different source domain combinations (S1+S3, S1+S3+S2, S1+S3+S2+S4) was compared with that of the fully integrated model on the target domain. (**c**) Four model strategies were compared: Naive Pooling, Weighted Ensemble, Naive Pooling + ATL, and Weighted Ensemble + ATL. (**d**) The importance of each source domain is evaluated by sequentially removing one domain at a time and comparing it with the fully integrated model.

We performed a systematic study to test the performance of our multi-source transfer learning framework. As Figure 2 shows, the model was tested with different amounts of target data and different source combinations. The results clearly prove that our multi-source approach is much better than using only a single source or simple data mixing. Even when the target data size is very small, such as 90 samples, the ensemble model maintains a high $R^2$ and a low MAE. This demonstrates that our framework can effectively extract useful physical information from many different systems to help with difficult 2D material tasks.

**Table 1**. Details of the source domains used in this study.

| Source ID | Source Name | Data volume |
|---|---|---|
| S1 | 3D Effective Mass[29] | 16569 |
| S2 | 3D Deformation Potential[30] | 7519 |
| S3 | 2D Formation Energy[31] | 7996 |

| S4 | 2D PBE Band Gap[31] | 3846 |
|---|---|---|
| S5 | 3D PBE Band Gap[9] | 16258 |

We compared how single sources perform against the Ensemble All strategy. As Figure 2a shows, the multi-source ensemble works better than every single-source model at all target data sizes from 90 to 180 samples. For example, S1 and S3 are the best single predictors, but their $R^2$ values are still much lower than the ensemble result. This shows that carrier mobility is a complex property that one single physical feature cannot fully explain. Instead, joining different knowledge from both 3D and 2D materials provides a more complete picture of the physics. The advantage of our ensemble approach is even more obvious when the target data is very small, which proves that multi-source learning is very helpful for solving data scarcity problems.

We also tested how the model improves as we add more source domains. Figure 2b shows that the $R^2$ increases and the MAE decreases as we grow the source pool from two domains to the full ensemble. The biggest jump in performance happens when we move from two sources to three by adding S2. This suggests that S2 provides unique physical information that is different from S1 and S3. Since deformation potential directly relates to how electrons scatter in a crystal, adding this source helps the model understand the transport physics much better. Usually, many machine learning models stop improving or even get worse when they use too many different types of data. However, our framework remains stable even when we add sources like S4 and S5 that are not closely related to the target. This proves that the alignment module effectively removes noise and the gating system ensures that every new data source adds positive value to the final prediction.

Moreover, a significant advantage of the framework lies in its exceptional data efficiency, particularly when target-domain samples are extremely limited. As illustrated in Figure. 2a&b, the multi-source integrated model achieved predictive accuracy at a target sample size of $N = 120$ that was comparable to, or in some cases exceeded, the performance of models trained on significantly larger target datasets. This phenomenon suggests that the integrated knowledge from property-related (S1, S3, S5) and structure-related (S2, S4) source domains provides a robust prior physical distribution. By borrowing high-fidelity physical descriptors from these diverse sources, the framework effectively compensates for the statistical sparsity of the target domain. Consequently, the model does not require an extensive volume of 2D mobility labels to converge; instead, it leverages the cross-domain invariant features to navigate the initial low-data discovery process. This capability is crucial for 2D material research, where the high computational cost of DFPT mobility calculations often precludes the acquisition of large-scale datasets.

To validate the methodological efficacy of our gating system and domain alignment components, we systematically benchmarked four algorithmic configurations: Naive Pooling, Weighted Ensemble, Naive Pooling + ATL, and the fully integrated Weighted Ensemble + ATL framework (Figure 2c). As illustrated in Figure 2c, a simple unweighted aggregation of source data (Naive Pooling) yields suboptimal performance under both standard transfer learning and adversarial alignment regimes. In contrast, the sample-adaptive weighted ensemble significantly minimizes the MAE while concurrently elevating the $R^2$ metric. This pronounced performance enhancement demonstrates that the adaptive weighting module successfully mitigates cross-domain noise by selectively identifying informative source features rather than assigning uniform domain source weights. Crucially, the synergistic combination of adversarial domain alignment and the adaptive ensemble achieves the highest predictive fidelity, confirming that global distribution alignment and localized sample weighting function as mutually reinforcing operations to maximize knowledge transfer.

We first compared our weighted ensemble strategy with a simple data mixing method (Naive Pooling). As shown in Figure 2c, our strategy yields superior performance in both traditional transfer learning and our alignment framework. Specifically, the weighted ensemble significantly reduces the MAE and improves the $R^2$ value. This proves that the model is smart enough to choose the best information instead of just treating all data sources as equal. The best performance is reached when we combine our alignment module with the weighted ensemble which shows that these two parts work together perfectly.

We also tested the importance of each source by removing them one by one. As Figure 2d shows, removing S1 and S3 causes the biggest drop in accuracy. This matches basic physics because effective mass is a main factor that decides how fast carriers move. The importance of formation energy suggests that the stability and the local environment of the atoms are also necessary for understanding 2D transport. While S2 and S5 have smaller effects, they still provide extra details like electronic screening that simple models might miss. Even if we remove the most important source (S1), the model does not fail and still gives reasonable predictions. This shows that our multi-source plan creates a back-up mechanism. Because different physical features are related to each other, the model can use other sources to fill the gap when one source is missing.

### 2.3. Model extrapolation capability

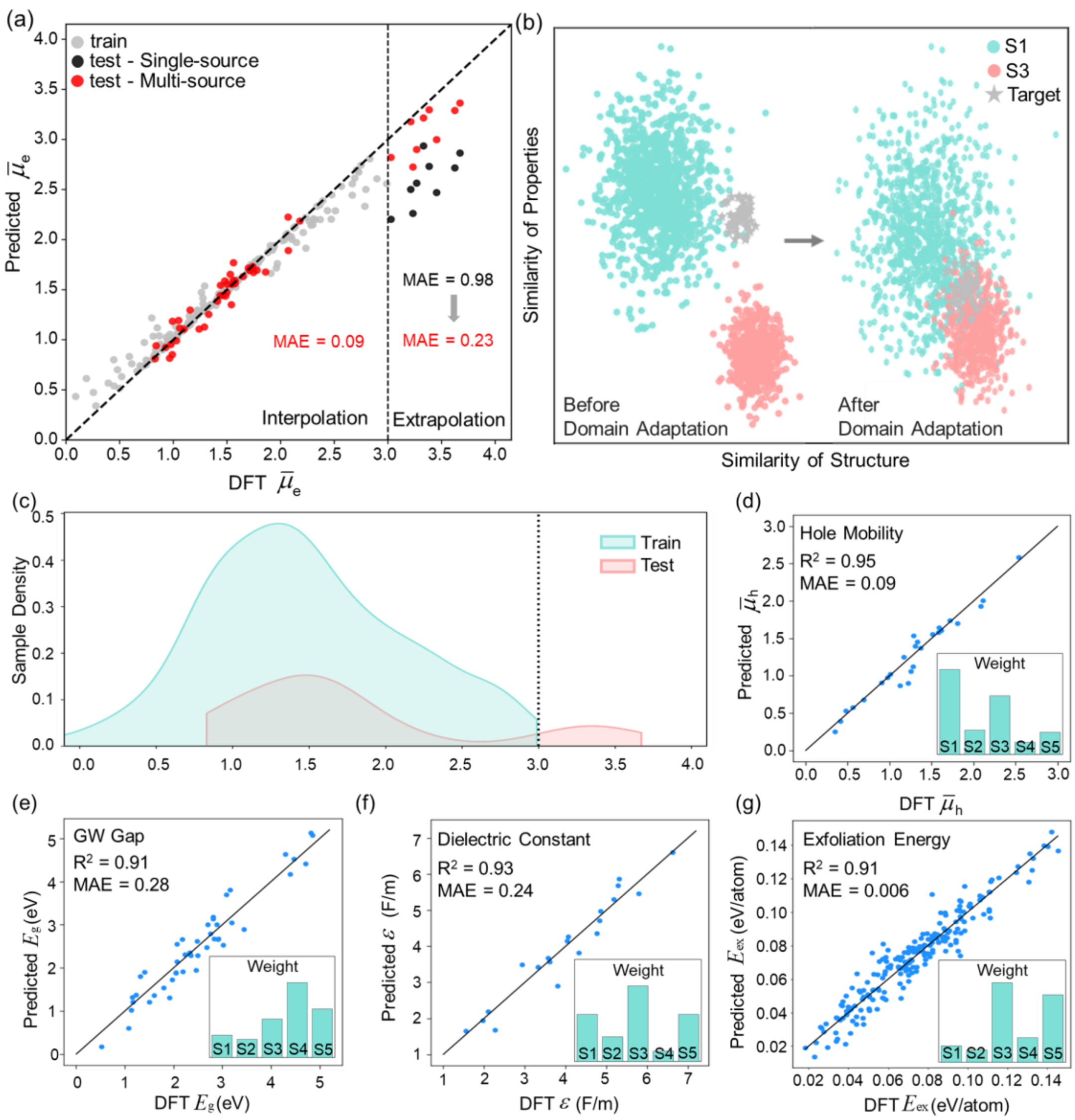

**Figure 3. Extrapolation performance of multi-Domain adversarial transfer learning.** (**a**) Parity plot of predicted vs. DFT average electron mobility. The train (gray), test-Single-source (black), and test-Multi-source (red) samples are plotted, with corresponding MAEs (0.09 for interpolation, 0.23 for extrapolation) labeled to show the model's generalization ability across interpolation/extrapolation intervals. (**b**) Feature space distribution before/after domain adaptation. S1 (cyan), S3 (pink), and Target (gray) samples are visualized; the arrow indicates that the target domain distribution aligns with source domains after adaptation, verifying the domain alignment effect. (**c**) Sample density distribution of average electron mobility for Train (light blue) and Test (pink) sets, showing the coverage of the test set (including extrapolation regions beyond the training distribution). (**d**)-(**g**) Testing sets for four material properties: parity diagrams and source domain weight distributions. (**d**) hole mobility, (**e**) GW gap, (**f**) dielectric constant, and (**g**) exfoliation energy. The blue scatter points represent the predicted vs. DFT values (with $R^2$ and MAE labeled to quantify prediction accuracy), and the green bar charts show the sample weight contribution of each source domain (S1-S5).

We also tested if our model works well for data it has never seen before and if it can predict properties other than electron mobility. As Figure 3 shows, the framework is flexible and stays accurate across different physical challenges. It not only handles extrapolation tasks but also adapts to a wide range of properties including electronic and thermodynamic features.

Predicting properties in areas where we have very little data is a main goal for discovering new materials. This is especially true for high-performance materials that sit outside the common data range. Figure 3a compares our predicted electron mobility with the real values from DFT. Our framework stays very accurate in the interpolation region where the error is only 0.09. More importantly, it shows a big improvement over single-source transfer models in the extrapolation region. While traditional single-source models fail badly with a high error of 0.98, our multi-source model keeps the error much lower at 0.23. The sample density statistics in Figure 3c further explain this improvement. It shows that combining data from many sources effectively expands the model's predictive horizon into sparse target regions. By using this multi-source information, the model can make accurate predictions even in the tail-distribution regions where target data is extremely scarce. This capability allows the framework to identify high-mobility materials that other models would completely miss.

We used dimensionality reduction in Figure 3b to show how the model improves its performance. Before domain adaptation, the features of the source domains (S1 and S3) and the target domain stay in separate groups. This large distance represents the domain gap that makes it difficult to transfer knowledge directly. After adversarial training, the feature points from these different domains move together and overlap significantly. This alignment proves that the framework effectively removes noise that is specific to one domain and finds universal physical features instead. We chose S1 and S3 for this analysis because they represent two different types of connections to the target. S1 has properties that are very similar to the target mobility. At the same time, S3 includes all the structural characteristics needed to describe the target 2D materials. By aligning these different sources, the model creates a unified physical space for more accurate predictions.

To prove that our framework learns universal physical rules rather than just memorizing one property, we challenged it with four other important material features. These include hole mobility and GW bandgap and dielectric constant and exfoliation energy. As shown in Figures 3d to 3g, the model maintains high accuracy across all these different tasks. The $R^2$ values range from 0.91 to 0.95 which confirms that the framework is very flexible. The small bar charts in each plot show that the model automatically changes its strategy for each property. For example, Figure 3e shows that S4 and S5 have the most influence when predicting the GW bandgap. This makes physical sense because these sources are closely related to how electrons behave in a crystal. On the other hand, S3 and S5 become more important for predicting exfoliation energy. This shows that the framework can prioritize stability-related knowledge when it deals with mechanical tasks. This ability to change weights based on the task proves that the model follows the complex rules of condensed matter physics. It shows that our method is a reliable tool for the wide 2D material landscape.

## 2.4. 2D semiconductors with high carrier mobility

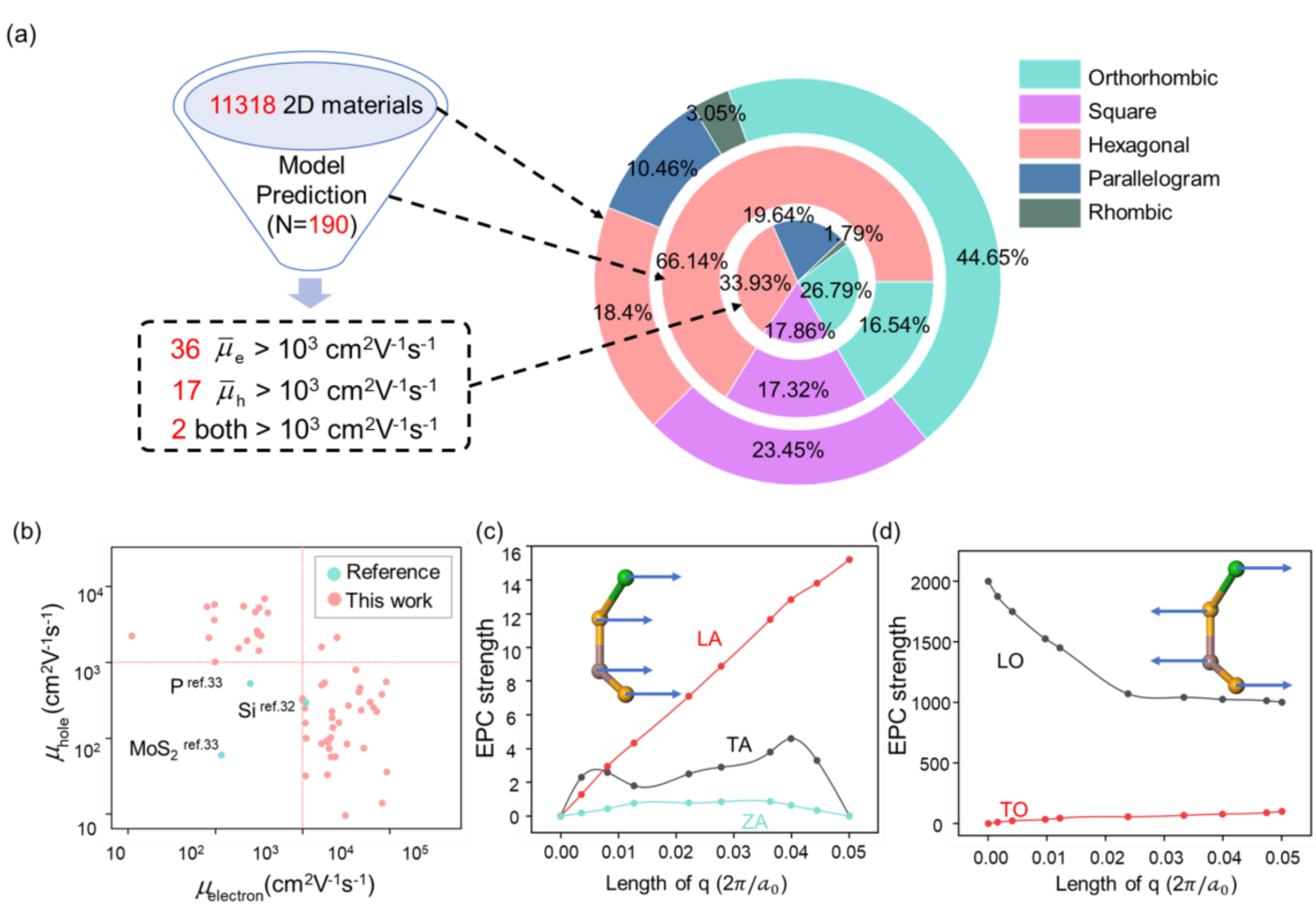

**Figure 4. Large-scale Screening and Discovery of High-Mobility 2D Materials.** (**a**) Screening results and structural evolution. Application of the model to 11,318 2D materials, identifying 55 candidates with mobility > $10^3$ cm$^2$V$^{-1}$s$^{-1}$ (including 36 exclusive n-type, 17 exclusive p-type, and 2 ambipolar candidate exhibiting both high electron and hole mobilities). (**b**) Electron and hole mobility of the ML screened 2D materials. Some common

semiconducting materials are also plotted as cyan dots for comparison[32,33]. (**c**–**d**) Electron-phonon coupling (EPC) in $BaIn_2Se_4$. EPC strength for (**c**) acoustic and (**d**) optical phonon modes. Initial states are 30 meV above CBM along Γ-*M*.

Building upon the robust predictive performance and generalization capabilities of the multi-source domain transfer learning framework, we applied the trained model to a large-scale screening task to identify novel high-performance 2D semiconductors. The final multi-source transfer learning model, trained on a optimized target-domain subset of 190 high-fidelity 2D carrier mobility samples, was utilized to scan a vast candidate pool consisting of 11,318 2D materials. Based on the model predictions, we successfully identified 55 promising candidates with exceptional transport properties. Compared to DFT-based mobility calculation, our approach is five orders of magnitude faster as shown in Supplementary Fig. 10. The transport landscape of the screened materials is visualized in Figure 4b, which plots hole mobility against electron mobility. The screened candidates occupy high-performance regions that significantly surpass common 2D semiconductors like $MoS_2$ and Si. Several newly discovered materials exhibit balanced high mobilities for both electrons and holes, comparable to or exceeding black phosphorus, making them ideal candidates for next-generation complementary metal-oxide-semiconductor (CMOS) logic devices. To verify the physical validity of the high-mobility predictions, we performed detailed electron-phonon coupling (EPC) analysis on a representative discovery, $BaIn_2Se_4$. Figures 4c and 4d illustrate the EPC "strength" $D_{mn\nu}(\boldsymbol{q}) = \sqrt{\omega_{\boldsymbol{q}\nu}}|g_{mn\nu}(\boldsymbol{q})|$[34,35] for selected phonon modes as a function of the scattering angle (here we use D instead of g to remove the common numerical instability in DFPT calculation for phonon frequency). The initial states were selected at 30 meV above the conduction band minimum (CBM) along the Γ-M direction, with final states located on the corresponding isoenergy circle. The longitudinal acoustic (LA) and transverse acoustic (TA) modes exhibit relatively low coupling strengths, while the longitudinal optical (LO) modes show a dominant influence on carrier scattering at certain angles (Fig. 4d). The relatively weak overall EPC strength in these identified modes explains the high intrinsic mobility predicted by the ML model, confirming that the framework successfully distills deep physical insights regarding lattice-vibration interactions without requiring explicit, computationally expensive EPC calculations for every candidate.

**Table 2**. Calculated carrier mobility for the top 10 materials with the highest carrier mobility

| **Material** | **Space Group** | **$\mu_e$($cm^2V^{-1}s^{-1}$)** | **$\mu_h$($cm^2V^{-1}s^{-1}$)** |
| --- | --- | --- | --- |
| $Ni_2Ru(Se_2I)_2$ | *Cm* | 6543 | 254 |
| $Co_2NiSe_3S$ | *P*3*m*1 | 9256 | 36 |
| $Ni_2GeS_2I$ | *Cm* | 5957 | 299 |
| $Li_2CrTe_3$ | *Amm*2 | 7102 | 225 |
| $CrGa_3Se_5I$ | *Cm* | 8138 | 380 |
| $BaIn_2Se_4$ | *P*3*m*1 | 2487 | 2128 |
| $Ba_2GaPtSe_3$ | *Pmm*2 | 8214 | 14 |
| $Li_2In_2PPdS_4$ | *P*4/*mmm* | 5038 | 74 |
| $Na_2TeAsIrSe_2$ | *Pmm*2 | 9095 | 555 |
| TiSeI | *Cm* | 367 | 6891 |
| $KNaInCuS_3$ | *Pmm*2 | 80 | 5421 |

Their crystal and electronic structures are given in Supplementary, and the structure files are provided in Supplementary Data 1.

## 3. Conclusion

In this study, we proved that the data-scarcity problem in 2D material research can be solved by learning from related crystal systems. Our framework successfully bridges the gap between bulk and 2D materials to find universal physical features. The accuracy of our model is much higher than traditional methods. For example, in unknown material regions, our multi-source approach reduced the prediction MAE from 0.98 to 0.23 compared to single-source models. This high precision allowed us to perform a large-scale screening of 11,318 2D materials. From this huge pool, we identified 55 promising candidates with carrier mobility higher than 1,000 $cm^2V^{-1}s^{-1}$. These results show that our model can effectively find high-performance materials even when the training data is extremely limited.

While our work marks a major step forward, it also provides a clear roadmap for future improvements. Currently, the framework still requires a small amount of high-fidelity EPC data to reach its best performance. To solve this, a very promising direction is to combine our multi-source model with Active Learning. This would allow the model to automatically choose the most important materials for new calculations, which would reduce the need for expensive data even further. Another exciting area for growth is the way we describe 2D structures[36]. Our current features are very efficient, but they do not fully capture how different layers interact with each other. Finally, although we focused on single-layer materials, this multi-source paradigm is a perfect fit for exploring van der Waals heterostructures. By training on these complex interfaces, the framework could help scientists design the next generation of electronic devices with customized transport properties. This study provides the foundation for using AI to navigate the most difficult areas of 2D material physics.

## 4. Methods

### 4.1 DFPT calculations

The first-principles calculations are performed using the Quantum Espresso (QE) Package[37] with Optimized Norm-Conserving Vanderbilt (ONCV) pseudopotentials[38,39] and the Perdew-Burke-Ernzerhof (PBE) exchange-correlation functional[40]. Specifically, the carrier mobility $\mu$ for band transport at low electric field can be obtained from the Boltzmann transport theory under momentum relaxation time approximation:

$$\boldsymbol{\mu_{\alpha\beta}} = \frac{q}{n_c\Omega}\sum\int \frac{dk}{\Omega_{BZ}}\frac{\partial f_{nk}}{\partial E_{nk}}\boldsymbol{\tau_{nk} v_{nk,\alpha} v_{nk,\beta}}, \tag{1}$$

where $\boldsymbol{\alpha}$ and $\boldsymbol{\beta}$ are the direction indices, $q$ is the charge of carrier, $\Omega$ ($\Omega_{BZ}$) is the area of unit cell (Brillouin zone); $\boldsymbol{v_{nk}}$ is its group velocity, and $E_{nk}$ is its energy; $f$ is the Fermi distribution function, and $n_c$ is the carrier density which is related with $f$ and the electronic band structure; $\boldsymbol{\tau_{nk}}$ is the momentum relaxation time for the electronic state with band index n and wavevector k, which can be calculated as:

$$\frac{1}{\boldsymbol{\tau_{nk}}} = \frac{2\pi}{\hbar}\sum\int \frac{dq}{\Omega_{BZ}}|\boldsymbol{g_{mnv}(k,q)}|^2[(\boldsymbol{f_{mk+q}} + \boldsymbol{n_{vq}})\boldsymbol{\delta}(\boldsymbol{E_{nk}} - \boldsymbol{E_{mk+q}} + \hbar\boldsymbol{\omega_{vq}}) + (1 + \boldsymbol{n_{vq}} - \boldsymbol{f_{mk+q}})\boldsymbol{\delta}(\boldsymbol{E_{nk}} - \boldsymbol{E_{mk+q}} + \hbar\boldsymbol{\omega_{vq}})](1 - \frac{\boldsymbol{v_{nk}} \cdot \boldsymbol{v_{mk+q}}}{|\boldsymbol{v_{nk}}||\boldsymbol{v_{mk+q}}|}) \tag{2}$$

where the initial electronic state $n_k$ is scattered to the final state $m_{k+q}$ by interacting with a phonon $\boldsymbol{v_q}$ with frequency $\omega_{vq}$ ($\nu$ is the phonon band index and q is the phonon wavevector); n is the Bose distribution; $\boldsymbol{v}$ is the group velocity vector; $\boldsymbol{g_{mnv}(k,q)}$ is the electron-phonon coupling (EPC) matrix element, which can be computed from density functional perturbation theory (DFPT)[41] and interpolated to a dense grid. More computational details can be found in the Supplementary Information (SI).

### 4.2 Data collection and processing

The target dataset consists of 230 mobility values for 2D materials, calculated via DFPT implemented in the QE. To ensure the highest data integrity, we moved beyond the conventional Deformation Potential Approximation (DPA)[42,43], which often oversimplifies scattering physics by neglecting non-polar phonon contributions and long-range coupling. Instead, The calculations of EPC and carrier mobility are performed using the EPW code[33], where the EPC are interpolated from sparse k/q grids to fine k/q grids. By accounting for these intricate many-body interactions, our dataset provides significantly higher fidelity than those derived from simplified analytical models. This high-fidelity foundation ensures that the physical descriptors distilled by the framework are grounded in the most accurate representation of 2D transport physics currently available. The Dielectric Constant is calculated from phonon calculations, while the GW bandgap originates is adopted from the paper[44], and the exfoliation energy data comes is adopted from the JARVIS[45] database.

To construct a comprehensive physical landscape, knowledge was integrated from five distinct source domains (S1–S5). Data for 3D effective mass (S1) and 3D bandgap (S5) were retrieved from the Materials Project[9] database, providing a robust bulk-phase baseline for electronic properties. 2D formation energy (S3) and 2D bandgap (S4) were sourced from the Computational 2D Materials Database (C2DB)[31,46], capturing the intrinsic stability and electronic characteristics of the 2D limit. S2 was extracted from a previously published high-throughput computational study focused on 2D structural and electronic screening . The elemental distribution and chemical diversity of these datasets are visualized in Supplementary Figures 1 and 2, confirming the broad coverage of the periodic table and the structural variety of the materials investigated.

### 4.3 Machine learning

The multi-source domain transfer learning framework is constructed with four primary components, a common feature extractor, a property regressor, a domain discriminator, and an adaptive weighting module. All neural network components are implemented using the PyTorch[47] framework. The feature extractor and property regressor utilize multi-layer perceptron (MLP) architectures, with hyperparameters—including the number of hidden layers and neurons per layer—optimized through a random search method to ensure robust latent representation. To extract domain-invariant features and bridge the gap between 3D bulk and 2D material domains, we implement an adversarial training strategy. During the training phase, the domain discriminator is trained to identify data sources, whereas the feature extractor is optimized to confuse the discriminator, thereby distilling universal physical representations. These distilled features are then processed by the adaptive weighting module, which dynamically assigns weights to multiple source domains (S1–S5) based on the inverse of their prediction errors. This mechanism prioritizes the most physically relevant source knowledge for each specific target sample, effectively mitigating the risk of negative transfer. The final high-dimensional feature vectors are fed into a gradient boosting tree (XGBoost) model to perform the final regression for carrier mobility and other properties. We also evaluated alternative regression algorithms, including kernel ridge regression (KRR) and least absolute shrinkage and selection operator (LASSO); however, the XGBoost framework consistently demonstrated superior performance in capturing the non-linear interdependencies of 2D transport physics.

## Acknowledgements

This work is supported by the National Key Research and Development Program of China (2022YFA1503103), Natural Science Foundation of China (T2321002, 22373013, 224B2303), Natural Science Foundation of Jiangsu Province, Major Project (BK20232012), Jiangsu Provincial Scientific Research Center of Applied Mathematics (BK20233002) and the Fundamental Research Funds for the Central Universities. The authors thank the computational resources from the Big Data Computing Center of SEU.

## Conflict of Interest

The authors declare no competing interests.